\documentclass[a4paper]{article}
\RequirePackage[english]{babel}
\RequirePackage[latin1]{inputenc}
\RequirePackage[T1]{fontenc}
\RequirePackage{mathrsfs}
\RequirePackage{amsmath}
\RequirePackage{amssymb}
\RequirePackage{amsbsy}
\RequirePackage{bm}
\begin{document}
\title{\bf{Conformal Gravity\\ with the most general ELKO Matter}}
\author{Luca Fabbri}
\date{}
\maketitle
\begin{abstract}
Recently we have constructed the conformal gravity with metric and torsion, finding the gravitational field equations that give the conservation laws and trace condition; in the present paper we apply this theory to the case of ELKO matter field, proving that their spin and energy densities once the matter field equations are considered imply the validity of the conservation laws and trace condition mentioned above.
\end{abstract}
\section*{Introduction}
That conformal gravity is important is due to a number of reasons: mathematically, its Lagrangian is unique as proven by Weyl, physically, its renormalizability was addressed by Stelle \cite{s}, phenomenologically, it provides an explanation for dark matter as discussed by Mannheim and Kazanas \cite{m-k}; spontaneous conformal symmetry breaking has been studied \cite{e-f-p/1,e-f-p/2}. On the other hand, a complete theory of gravity possesses beside the curvature also the torsion tensor, arising as gauge strengths of rotations and translations in the gauge theory of the Poincar\'{e} group \cite{h-h-k-n}; in this way the full coupling to both energy and spin density tensors may be established \cite{f}. Therefore, we have that metric as well as torsion conformal transformations have to be defined \cite{sh}. In a very recent paper, we have found a curvature with metric and torsion that is conformal in $(1+3)$-dimensional spacetimes, constructing the conformal metric-torsional theory by giving the system of gravitational field equations and the corresponding conservation laws and trace condition for the spin and energy densities \cite{F,FABBRI}.

Now, a few years ago, a new form of matter was introduced, and called ELKO from the acronym of the German \textit{Eigenspinoren des LadungsKonjugationsOperators}, designating eigenspinors of the charge-conjugation operator; ELKO fields are spin-$\frac{1}{2}$ fermions of Majorana type: since ELKOs are topologically neutral, they display non-locality \cite{a-g/1,a-g/2}. In order for ELKO fields to possess mass they have to obey second-order derivative matter field equations, meaning that their mass dimension is $1$ \cite{a-l-s/1,a-l-s/2}. That ELKO are Majorana fields with scalar-like mass dimension endows them with interesting properties, and their possible applications ranging from particle physics to cosmology have been addressed in various works such as those given in literature in references from \cite{f/1} to \cite{fabbri-v}.

In this paper, we shall apply the above-mentioned theory of conformal gravity with curvature and torsion to the case of ELKO matter fields: after a brief introduction of the geometrical background we will discuss how to construct an ELKO action that is conformally invariant; its variation will give the ELKO spin and energy densities with matter field equations, with which we will check that both conservation laws and trace condition are valid indeed. Eventually we will discuss how to ensure the conformal invariance then both curvature and torsion tensors are in fact necessary in the case of the ELKO matter model.
\section{Conformal Curvature and Derivative}
Here we follow the convention about geometry as in \cite{F}, and for matter as in \cite{FABBRI}.

We only recall that for a given $\sigma$ and defining $\ln{\sigma}=\phi$ the torsion and metric tensors have conformal transformations
\begin{eqnarray}
&Q^{\sigma}_{\phantom{\sigma}\rho\alpha}\rightarrow Q^{\sigma}_{\phantom{\sigma}\rho\alpha}
+q(\delta^{\sigma}_{\rho}\partial_{\alpha}\phi-\delta^{\sigma}_{\alpha}\partial_{\rho}\phi)
\label{torsion}\\
&g_{\alpha\beta}\rightarrow\sigma^{2}g_{\alpha\beta}
\label{metric}
\end{eqnarray}
so that for the metric-compatible connection $\Gamma^{\sigma}_{\phantom{\sigma}\rho\alpha}$ the conformal transformation is a consequence of (\ref{torsion}) and (\ref{metric}); in terms of metric and connection it is possible to introduce the Minkowskian metric $\eta_{ij}$ and a basis of vierbein $e_{\alpha}^{i}$ together with the antisymmetric spin-connection $\omega^{ij}_{\phantom{ij}\alpha}$ and their conformal transformation is given in terms of the previous ones: eventually we introduce the $\boldsymbol{\gamma}_{a}$ matrices so to define the $\frac{1}{4}[\boldsymbol{\gamma}_{a},\boldsymbol{\gamma}_{b}]=\boldsymbol{\sigma}_{ab}$ matrices through which the connection $\boldsymbol{\Omega}_{\rho}$ and its conformal transformation are assigned, and covariant derivatives $\boldsymbol{D}_{\rho}$ for ELKO fields may be defined. The ELKO and ELKO dual conformal transformation is
\begin{eqnarray}
&\lambda\rightarrow\sigma^{-1}\lambda\ \ \ \ \ \ \ \ \ \ \ \ \ \ \ \ \stackrel{\neg}{\lambda}\rightarrow\sigma^{-1}\stackrel{\neg}{\lambda}
\label{ELKO}
\end{eqnarray}
as ELKO and ELKO dual are Majorana fields with scalar-like mass dimension.

The Riemann-Cartan metric-torsional curvature tensor $G_{\rho\xi\mu\nu}$ is defined as it is customary. The commutator of spinorial covariant derivatives is
\begin{eqnarray}
&[\boldsymbol{D}_{\rho},\boldsymbol{D}_{\mu}]\lambda
=Q^{\theta}_{\phantom{\theta}\rho\mu}\boldsymbol{D}_{\theta}\lambda
+\frac{1}{2}G^{\alpha\beta}_{\phantom{\alpha\beta}\rho\mu}\boldsymbol{\sigma}_{\alpha\beta}\lambda
\label{conformalcommutator}
\end{eqnarray}
and once the curvature is defined, this is a geometric identity.

The metric-torsional curvature tensor can be modified as in the following
\begin{eqnarray}
&M_{\alpha\beta\mu\nu}
=G_{\alpha\beta\mu\nu}+(\frac{1-q}{3q})(Q_{\beta}Q_{\alpha\mu\nu}-Q_{\alpha}Q_{\beta\mu\nu})
\label{curvature}
\end{eqnarray}
since this is the form whose irreducible part
\begin{eqnarray}
&T_{\alpha\beta\mu\nu}=M_{\alpha\beta\mu\nu}
-\frac{1}{2}(M_{\alpha[\mu}g_{\nu]\beta}-M_{\beta[\mu}g_{\nu]\alpha})
+\frac{1}{12}M(g_{\alpha[\mu}g_{\nu]\beta}-g_{\beta[\mu}g_{\nu]\alpha})
\label{conformalcurvature}
\end{eqnarray}
is conformally covariant in $(1+3)$-dimensional spacetimes. However, because the ELKO field is a Majorana spinor with a peculiar mass dimension, from the covariant derivative of ELKO the construction of a kinetic term that is also conformally invariant is not a straightforward procedure, as we discuss next.
\section{Conformal Gravity and Matter: ELKO Fields}
When treating the conformal spinor field, the simplest example of Dirac field is such that due to its mass dimension $\frac{3}{2}$ its first-order derivative kinetic term and conformal transformations makes it conformally invariant in a trivial way \cite{FABBRI}.

The same cannot hold in this case, since ELKO are Majorana fields with mass dimension $1$: although they a special type of spinor, their peculiar mass dimension makes them more analogous to scalars both about their dynamical properties and in terms of their conformal transformation; so for ELKO like scalar fields, the second-order derivative kinetic term $|\boldsymbol{D}\lambda|^{2}$ is such that the conformal transformation (\ref{ELKO}) produces the appearance of additional extra terms in the action. However, we may take advantage of the fact that geometrical fields like the torsion trace $Q_{\alpha}$ and the modified curvature trace $M$ under their conformal transformations produce the appearance of similar types of additional terms in the action: therefore for ELKO as well as for scalars, if beside the kinetic term also these extra terms are added, then specific fine-tunings may be assumed for which all additional terms cancel exactly, yielding for this improved kinetic term in the action the conformal invariance. A quick inventory gives possible terms that can be taken and a long although straightforward calculation shows that a specific fine-tuning may actually be chosen in order for the action to be conformally invariant if given by
\begin{eqnarray}
\nonumber
&S=\int[L_{\mathrm{gravity}}
+\boldsymbol{D}_{\rho}\stackrel{\neg}{\lambda}\boldsymbol{D}^{\rho}\lambda
+a\boldsymbol{D}_{\rho}\stackrel{\neg}{\lambda} \boldsymbol{\sigma}^{\rho\beta}\boldsymbol{D}_{\beta}\lambda+\\
\nonumber
&+(\frac{aq+q-1}{3q})
(\boldsymbol{D}_{\alpha}\stackrel{\neg}{\lambda}\boldsymbol{\sigma}^{\alpha\nu}\lambda
-\stackrel{\neg}{\lambda}\boldsymbol{\sigma}^{\alpha\nu}\boldsymbol{D}_{\alpha}\lambda)Q_{\nu}
+(\frac{4-3a+3qa-24p+24pq}{12q})\boldsymbol{D}_{\nu}\lambda^{2}Q^{\nu}+\\
&+(\frac{7-3a-6q+3aq^{2}+3q^{2}-24p-24pq+48pq^{2}}{36q^{2}})\lambda^{2}Q_{\alpha}Q^{\alpha}
+p\lambda^{2}M]|e|dV
\label{action}
\end{eqnarray}
with parameters $a$ and $p$ and where it is over the volume of the spacetime that the integral is taken. By varying (\ref{action}) we get the spin and energy densities
\begin{eqnarray}
\nonumber
&S^{\mu\alpha\beta}=
\frac{1}{2}(\boldsymbol{D}^{\mu}\stackrel{\neg}{\lambda}\boldsymbol{\sigma}^{\alpha\beta}\lambda
-\stackrel{\neg}{\lambda}\boldsymbol{\sigma}^{\alpha\beta}\boldsymbol{D}^{\mu}\lambda)+\\
\nonumber
&+\frac{a}{2}(\boldsymbol{D}_{\rho}\stackrel{\neg}{\lambda}
\boldsymbol{\sigma}^{\rho\mu}\boldsymbol{\sigma}^{\alpha\beta}\lambda
-\stackrel{\neg}{\lambda}\boldsymbol{\sigma}^{\alpha\beta}\boldsymbol{\sigma}^{\mu\rho}
\boldsymbol{D}_{\rho}\lambda)+\\
\nonumber
&+(\frac{1-q-aq}{6q})(\boldsymbol{D}^{\rho}\stackrel{\neg}{\lambda}
\boldsymbol{\sigma}_{\rho\theta}\lambda
-\stackrel{\neg}{\lambda}\boldsymbol{\sigma}_{\rho\theta}
\boldsymbol{D}^{\rho}\lambda)g^{\theta[\beta}g^{\alpha]\mu}-\\
\nonumber
&-(\frac{4-3a+3aq-24p}{24q})\boldsymbol{D}_{\theta}\lambda^{2}g^{\theta[\beta}g^{\alpha]\mu}
-(\frac{1-q-aq}{6q})Q_{\theta}
\stackrel{\neg}{\lambda}\{\boldsymbol{\sigma}^{\theta\mu},\boldsymbol{\sigma}^{\alpha\beta}\}\lambda+\\
&+(\frac{7-3a-6q+3q^{2}+3aq^{2}-22p-29pq+48pq^{2}}{36q^{2}})Q^{[\alpha}g^{\beta]\mu}\lambda^{2}
-pQ^{\mu\alpha\beta}\lambda^{2}
\label{spin}\\
\nonumber
&T^{\alpha\mu}=
(\boldsymbol{D}^{\mu}\stackrel{\neg}{\lambda}\boldsymbol{D}^{\alpha}\lambda
+\boldsymbol{D}^{\alpha}\stackrel{\neg}{\lambda}\boldsymbol{D}^{\mu}\lambda
-g^{\mu\alpha}\boldsymbol{D}_{\rho}\stackrel{\neg}{\lambda}\boldsymbol{D}^{\rho}\lambda)+\\
\nonumber
&+a(\boldsymbol{D}^{\mu}\stackrel{\neg}{\lambda}
\boldsymbol{\sigma}^{\alpha\rho}\boldsymbol{D}_{\rho}\lambda
+\boldsymbol{D}_{\rho}\stackrel{\neg}{\lambda}
\boldsymbol{\sigma}^{\rho\alpha}\boldsymbol{D}^{\mu}\lambda
-g^{\mu\alpha}\boldsymbol{D}_{\rho}\stackrel{\neg}{\lambda}
\boldsymbol{\sigma}^{\rho\sigma}\boldsymbol{D}_{\sigma}\lambda)+\\
\nonumber
&+(\frac{1-q-aq}{3q})\boldsymbol{D}_{\beta}(\boldsymbol{D}^{\rho}\stackrel{\neg}{\lambda} \boldsymbol{\sigma}_{\rho\nu}\lambda
-\stackrel{\neg}{\lambda}\boldsymbol{\sigma}_{\rho\nu}\boldsymbol{D}^{\rho}\lambda)
(g^{\alpha\nu}g^{\mu\beta}-g^{\alpha\mu}g^{\nu\beta})-\\
\nonumber
&-(\frac{4-3a+3qa-24p+24pq}{12q})\boldsymbol{D}_{\beta}\boldsymbol{D}_{\nu}\lambda^{2}
(g^{\alpha\nu}g^{\mu\beta}-g^{\alpha\mu}g^{\nu\beta})+\\
\nonumber
&+(\frac{1-q-aq}{3q})Q_{\rho}(\boldsymbol{D}^{\mu}\stackrel{\neg}{\lambda}
\boldsymbol{\sigma}^{\rho\alpha}\lambda
-\stackrel{\neg}{\lambda}\boldsymbol{\sigma}^{\rho\alpha}\boldsymbol{D}^{\mu}\lambda)-\\
\nonumber
&-(\frac{7-3a-6q+3q^{2}+3aq^{2}-24p+24pq^{2}}{18q^{2}})\boldsymbol{D}_{\beta}(\lambda^{2}Q_{\nu})
(g^{\beta\mu}g^{\alpha\nu}-g^{\alpha\mu}g^{\beta\nu})+\\
\nonumber
&+(\frac{4-3a+3qa-24p+24pq}{12q})Q^{\alpha}\boldsymbol{D}^{\mu}\lambda^{2}+\\
\nonumber
&+(\frac{7-3a-6q+3q^{2}+3aq^{2}-24p+24pq}{36q^{2}})Q_{\rho}Q^{\rho}\lambda^{2}g^{\alpha\mu}-\\
&-\frac{2p(1-q)}{3q}(Q^{\alpha}Q^{\mu}+Q^{\alpha\mu\rho}Q_{\rho})\lambda^{2}
+2p(M^{\alpha\mu}-\frac{1}{2}g^{\alpha\mu}M)\lambda^{2}
\label{energy}
\end{eqnarray}
along with the matter field equations
\begin{eqnarray}
\nonumber
&\boldsymbol{D}^{2}\lambda+Q^{\mu}\boldsymbol{D}_{\mu}\lambda
+a\boldsymbol{\sigma}^{\rho\mu}\boldsymbol{D}_{\rho}\boldsymbol{D}_{\mu}\lambda
+(\frac{aq-2q+2}{3q})Q_{\rho}\boldsymbol{\sigma}^{\rho\mu}\boldsymbol{D}_{\mu}\lambda+\\
\nonumber
&+(\frac{aq+q-1}{3q})D_{\rho}Q_{\mu}\boldsymbol{\sigma}^{\rho\mu}\lambda
+(\frac{4-3a+3qa-24p+24pq}{12q})D_{\rho}Q^{\rho}\lambda-\\
&-(\frac{7-3a-18q+9aq-6aq^{2}+3q^{2}-24p+48pq-24pq^{2}}{36q^{2}})Q_{\alpha}Q^{\alpha}\lambda
-pM\lambda=0
\label{matterequations}
\end{eqnarray}
as a simple although quite laborious computation would show.

It is possible to see that the spin and energy densities (\ref{spin}-\ref{energy}) so soon as the conformal matter field equations (\ref{matterequations}) are accounted have conservation laws
\begin{eqnarray}
&D_{\rho}S^{\rho\mu\nu}+Q_{\rho}S^{\rho\mu\nu}
+\frac{1}{2}T^{[\mu\nu]}=0
\label{conservationlawspin}\\
&D_{\mu}T^{\mu\rho}+Q_{\mu}T^{\mu\rho}-T_{\mu\sigma}Q^{\sigma\mu\rho}
+S_{\beta\mu\sigma}G^{\sigma\mu\beta\rho}=0
\label{conservationlawenergy}
\end{eqnarray}
and the trace condition
\begin{eqnarray}
&(1-q)(D_{\mu}S_{\nu}^{\phantom{\nu}\nu\mu}+Q_{\mu}S_{\nu}^{\phantom{\nu}\nu\mu})
+\frac{1}{2}T_{\mu}^{\phantom{\mu}\mu}=0
\label{trace}
\end{eqnarray}
where the commutator of covariant derivatives has been used; when a geometrical background is conformally invariant there is the loss of one degree of freedom realized by the introduction of the trace condition as constraint: such constraint has the structure of a conservation law and it is therefore dynamically implemented in the conformal theory of gravitation for ELKO matter fields.
\section*{Conclusion}
We have considered the conformal theory of gravity discussing the case of ELKO matter fields, of which we have obtained the spin and energy density tensors along with the matter field equations, showing that the spin and energy densities, when the matter field equations are taken, satisfy both conservation laws and trace condition; we have to stress that among all possible values of the conformal parameters $q$, $a$, $p$, there exists no possible choice for which the presence of torsion can be removed altogether: if a conformal gravity that beside curvature has also torsion is important in itself, a conformal curvature-torsional gravitational theory is necessary for applications to the ELKO matter model.

\end{document}